\documentclass[12pt,letterpaper]{article}
\usepackage{jheppub}

\usepackage{graphicx}
\usepackage{bbm}
\usepackage{amsmath}
\usepackage{amssymb}
\usepackage{mathtools}
\usepackage{amsthm}

\usepackage[labelformat=simple]{subcaption}

\def\bea{\begin{eqnarray}}
\def\eea{\end{eqnarray}} 
\def\be{\begin{equation}}
\def\ee{\end{equation}}
\newcommand{\ba}{\begin{eqnarray}}
\newcommand{\ea}{\end{eqnarray}}

\def\lcz{\lambda^{(0)}_c }
\def\kb{\kappa_b}
\def\kc{\kappa_c}
\def\ks{\kappa_\text{sys}}

\def\rbh{r_b }
\def\rds{r_c }

\def\qt{ \tilde{a} }
\def\at{ \tilde{a} }

\def\be{\begin{equation}}
\def\ee{\end{equation}}

\begin{document}

\begin{titlepage}

\setcounter{page}{1} \baselineskip=15.5pt \thispagestyle{empty}

\bigskip\

 \vspace{2cm}
\begin{center}
{\Large \bfseries  Constrained Spin Systems and KNdS Black Holes}
\end{center}
 
\vspace{1cm}

\begin{center}
\scalebox{0.95}[0.95]{{\fontsize{14}{30}\selectfont Vivek Chakrabhavi$^{a,b}$, Muldrow Etheredge$^a$, Yue Qiu$^{a,c}$, and Jennie Traschen$^a$ 
}}
\end{center}

\begin{center}
\vspace{0.25 cm}
\textsl{$^{a}$Department of Physics, University of Massachusetts, Amherst, MA 01003, USA}\\
\textsl{$^{b}$Department of Physics and Astronomy, University of Pennsylvania, Philadelphia, PA 19104, USA} \\
\textsl{$^{c}$Department of Physics, Tsinghua University, Beijing 100084, China}\\
\vspace{0.25cm}
\end{center}

\vskip 0.1 in Email: \texttt{vivekcm@sas.upenn.edu, metheredge@umass.edu,  yqiu@umass.edu, 
\\traschen@umass.edu}
\vspace{6pt}
\vskip 0.1in
\par

\begin{center}
{\bf Abstract}  
\end{center}
\begin{quote}

Kerr-Newman de Sitter (KNdS) spacetimes have a rich thermodynamic structure that involves multiple horizons, and so differs in key
respects from asymptotically flat or AdS black holes.
 In this paper, we show that certain features of KNdS spacetimes can be reproduced by a constrained system of $N$ non-interacting spins in a magnetic field. Both the KNdS and spin systems have bounded energy and entropy, a maximum of the entropy in the interior of the energy range, and
 a symmetry that maps lower energy states to higher energy states with the same entropy.
 Consequently, both systems have a temperature that can be positive or negative, where the gravitational temperature is defined
 analogously to that of the spins. We find that the number of spins $N$ corresponds to $1/\Lambda$ for black holes with very small charge $q$ and rotation parameter $a$, and scales like $\sqrt{(a^2+q^2)/\Lambda}$ for larger values of $a$ and $q$. By studying constrained spin systems, we provide insight into the thermodynamics of KNdS spacetimes and its quantum mechanical description.

\vfill

\vskip 2.mm
\end{quote}
\end{titlepage}

\setcounter{tocdepth}{2}

\hrule
\tableofcontents

\bigskip\medskip
\hrule
\bigskip\bigskip

\section{Introduction}
Rotating, charged, black holes in de Sitter, or Kerr-Newman de Sitter black holes (KNdS)
 have thermodynamic properties that differ from  asymptotically flat or AdS black holes in significant ways.
 When $\Lambda$ is zero or negative, the mass $M$ of the black hole and its horizon area 
$A_b$ can be
arbitrarily large. However, when $\Lambda$ is positive a cosmological horizon surrounds
the black hole horizon, leading to an upper bound on $M$ and $A_b$.  The black hole surface gravity $\kb$, or temperature, is positive 
in all these cases, but the surface 
gravity of the cosmological horizon $\kc$ is negative. Interestingly, in KNdS spacetimes the total horizon area, identified as the gravitational entropy,
 does not simply increase with $M$.
Rather,  $A=A_b +A_c$ is maximized\footnote{In this paper, we hold the cosmological constant $\Lambda$ fixed and positive.} at an intermediate mass. This leads to the unusual situation
 in which the thermodynamic temperature derived from $A(M)$ changes from positive to negative. However, 
these unusual features of KNdS black holes are also core properties of the thermodynamics of a finite set of  spins in a magnetic field. 
Exploring this commonality is the subject of this paper.

The idea that a quantum description of black holes should be based on a finite dimensional Hilbert space has been the subject of much 
research and conjecture.  The finite entropy of de Sitter \cite{Gibbons:1977mu} has prompted the idea that the Hilbert space for de Sitter spacetimes is finite dimensional \cite{Banks:2000fe, Witten:2001kn, Banks:2001yp, Arenas-Henriquez:2022pyh, Arias:2019pzy, Banks:2018ypk, Banks:2020zcr},
and has motivated construction of quantum models with a finite number of degrees of freedom, including matrix models
\cite{Banks:2005bm, Kabat:2002hj, Banks:2003cg, Parikh:2004wh, Banks:2006rx, Anninos:2022ujl, Li:2001ky, Susskind:2021dfc} 
and SYK models in the high-temperature limit
\cite{Susskind:2021esx, Susskind:2022bia, Rahman:2022jsf, Goel:2023svz}.
Spin systems are commonly used to model the scrambling of information as well as
black hole radiation \cite{Corley:1997ef, Lowe:2015eba, Lowe:2017ehz, Lowe:2019scv}. Additionally, references \cite{Dinsmore:2019elr, Johnson:2019vqf, Johnson:2019ayc} investigate Schottky anomalies\footnote{A Schottky anomaly here refers to a situation where the energy and entropy as a function of temperature  have a peak.}.
The causal set formulation of quantum gravity is by construction locally finite and has been applied to 
black hole entropy \cite{Dou:2003af, Barton:2019okw}.

In this paper we compare the thermodynamics of a KNdS black hole to that of a system of spins in a magnetic field, and we present evidence 
that a Hilbert space for a KNdS black hole has many common features with that of the spin system. The comparison we present is stripped-down
as the spins are assumed to be non-interacting. It is interesting to
see how much relevant structure emerges nonetheless. 

A system of $N$ non-interacting spins in a magnetic field is a simple model for a paramagnet. The energy of the system $E$ is bounded,
and unlike many familiar thermodynamic systems,
the entropy $S(E)$ does not simply increase with $E$. Instead the maximum of $S$ occurs at an intermediate value of energy, and
$S$  is symmetric upon interchanging the number of  spins ``up" and the number of spins ``down". The energy does distinguish between
spin up and spin down,
 and $E$ increases monotonically from the state with all spins up to the state with all down.
As a result, the temperature of the system is positive on the low $E$ side of the $S(E)$ curve, negative on the other, and goes to 
infinity where the entropy is maximized.

The total area $A =A_b +A_c$ of the black hole plus cosmological horizons turns out to have
 a symmetry that maps between distinct spacetimes having the same total area but different masses \cite{McInerney:2015xwa}. 
Hence $A(M)$ is maximized in the middle of the mass range, as with the entropy of the spin system. The derivative of the
$A(M)$ curve defines a ``surface gravity of the system" $\kappa_\text{sys}$ which is positive on the lower $M$ side of the curve, negative on the 
large $M$ side, and goes to infinity where the total area is maximized.  This is the same behavior as the temperature of the spin system,
compare Figures \ref{f.AofM} and \ref{f.ksysofy} to Figure \ref{f.spinfig}.

The two first laws of thermodynamics for KNdS black holes yield a simple expression for $\kappa_\text{sys}$ in terms of $\kb$ and $\kc$, 
 showing that
 negative values of $\ks$ follow from the fact that $\kappa_c$ is negative. The infinity in $\ks$ occurs when the two surface
gravities sum to zero, \emph{i.e.}, when they have equal magnitude.
We observe that, rather than the negative value of $\kc$ being a pesky minus sign,
one can understand it as a consequence of a system with (1) a bounded energy, and (2) a symmetry of the entropy that maps between
distinct isentropic spacetimes.

KNdS spacetimes naturally have distinct $``b"$ and $``c"$ features. 
To see whether the spin model can also contain these features (following \cite{Dinsmore:2019elr}) we divide the spins into
constrained $``b"$ and $``c"$ subsystems, an approach which is partially successful and motivates
educated speculation regarding what spins up or spins down correspond to in the black hole spacetimes. 
The correspondence between area and entropy is used to derive an exact expression (within the model) for
 how the total number of spins $N$ depends on the parameters of the
KNdS black holes. We find that when the black hole area is negligible compared to that of the cosmological horizon, then
\begin{align}
	N\sim  1 /(\hbar G^2  \Lambda ).
\end{align}
When the areas of the two horizons are parametrically comparable, we find
\begin{align}
	N\sim  \sqrt{ a^ 2 + q^2} / (\hbar G^{3/2} \Lambda^{1/2} ) ,
\end{align}
where $q$ is the charge and $a$ is the rotation parameter of the 
 black hole.

\section{Review of KNdS back hole solutions}

In this section we review the KNdS metric and give expressions for the thermodynamic quantities that are used in the paper.
Some of the formulas are explained in more detail in Appendix \ref{appendix}.  Much of this follows previous work of \cite{McInerney:2015xwa},
which introduces a new and useful parameterization of KNdS spacetimes and analyzes the black hole parameter space. 

After reviewing \cite{McInerney:2015xwa}, we define a new parameter $y$ which facilitates computing the ``system surface gravity". This parameter $y$ makes manifest a symmetry
of the total horizon area in a way that corresponds with the symmetry of the entropy of the spin system described in Section \ref{s.SpinSystems}.

\subsection{Kerr-Newman-deSitter black hole solutions}

In $D=4$ the KNdS metric for a rotating, charged black hole with positive cosmological constant $\Lambda =3/l^2 $ is \cite{Carter:1973rla}
	\begin{align}
	\begin{aligned}
		\label{rotmetric}
ds^2   &=    -\frac{\Delta}{\rho^2}\left(dt-\frac{a\sin^2\!\theta}{\gamma} d\varphi\right)^2 +\frac{\rho^2}{\Delta} dr^2
+\frac{\rho^2}{\Psi} d\theta^2 \\ \cr 
& +
\left(1+{a^2 \over l^2 } \cos^2\!\theta \right) \frac{\sin^2\!\theta}{\rho^2}\left(adt-\frac{r^2+a^2}{\gamma}d\varphi\right)^2,
\end{aligned}
	\end{align}
where
\begin{align}
\begin{aligned}
	\label{delta}
 \quad  \Delta &=\left (r^2+a^2\right )\left (1- {r^2\over l^2 } \right )-2mr+ q^2, & \Psi &=1+{a^2 \over l^2 } \cos^2\!\theta  \,, \\
  \rho^2 &= r^2+a^2\cos^2\!\theta\,,&\quad  \gamma &=1+a^2 / l^2.
\end{aligned}
\end{align}
The gauge field is given by $A = -\frac{q r}{ \rho^2}  (dt\!- { a\over \gamma} \sin^2 \theta  \ d\varphi  )$. The ADM mass $M$,  angular momentum $J$, and electric charge $Q$  are related to the metric parameters $m, a$ and $q$ by
\be\label{ADMcharges}
M=m / \gamma^2 \  , \  \ J= am / \gamma^2  \ , \ \ Q= q /\gamma.
\ee
Note that $\gamma$ is a constant when $a$ and $l$ are kept fixed,  but that $\gamma $ is
not constant  for fixed $J$ and $l $.  
The angular velocities and the electric potential at each horizon are given by\footnote{Details can be found in \cite{Mellor:1989wc}, and an analysis  of the global spacetime structure is given in \cite{Akcay:2010vt}. }
\begin{align}
	 \Omega_{h}= a \gamma/ (r_{h}^2+a^2 ) , \qquad   \phi_{h} = q  r_{h} / (r_{h}^2+a^2 ),
\end{align}

The horizons  occur at the roots  $r_h$  of $\Delta$ in \eqref{delta} where $\Delta (r_h ) = 0$ and  $r_I < r_b < r_c $ denoting the inner Cauchy, the black hole,
and the cosmological horizons. There is also a negative root.
The areas of the black hole and the cosmological horizons are given by 
\begin{equation}\label{bhdsav}
 A_{h} = {4\pi \over \gamma} (r_{h}^2+a^2) , \qquad h\in \{b,c\}.
 \end{equation}
The surface gravity at each horizon is given by
\begin{align}
	2 \kappa_h = (r_h^2 +a ^2) ^{-1} \Delta^\prime (r_h ).
\end{align} 
The explicit formulae are
given in (\ref{kndstemps}). 
The surface gravity is positive at the black hole horizon and negative at the deSitter horizon.
The fact that the de Sitter surface gravity is negative because $\Delta (r)$ is positive
between the two adjacent roots $r_b < r_c$ so that
$t $ a timelike coordinate.
The surface gravity $\kappa_c$ is negative, so one might prefer to work with the positive quantity $2\pi T_c = - \kappa_c$. However, we will not do this in this paper, because in the comparison of
KNdS and a spin system below, negative temperature occurs naturally in part of the spin phase space, and this corresponds to 
the negative cosmological surface gravity.

We next examine specific properties of the total area and the surface gravities which are key for comparison to the spin system.
Additional previous studies of the classical and quantum thermodynamics of KNdS include \cite{Davies:1989ey, Romans:1991nq, Dehghani:2002nt, Gibbons:2004uw, Ghezelbash:2004af, Sekiwa:2006qj, Dolan:2013ft, Altamirano:2013ane, Ma:2013aqa, Zhao:2014zea, Kubiznak:2015bya, Bhattacharya:2015mja, Li:2016zdi, Hajian:2016kxx, Kubiznak:2016qmn, Pappas:2017kam, Bhattacharya:2017scw, Gregory:2017sor, Cvetic:2018dqf, Du:2022jcb, Zhen:2022bsy, Li:2021axp, Ali:2019rjn, Gregory:2018ghc, Banihashemi:2022htw, Bhattacharya:2018ltm, Qiu:2019qgp, Gregory:2021ozs}. 
To keep the notation distinct, we will refer to surface gravity and area
in the black hole analysis, and use the terms entropy and temperature for the spin systems, though at some instances it is useful to
use the terminology ``gravitational temperature".

\subsection{Symmetry of the area product parameter $X=r_br_c$}

The metric contains four parameters $l$, $a$, $q$, and $m$. But, the areas and surface gravities depend these parameters through $r_h$, and
each $r_h$ is the solution to a quartic equation. This situation is not immediately so convenient for analysis, because many objects of thermodynamical interest in the previous subsection are expressed as functions of $r_h$. To deal with this, we consider the $X$ parameter of
 \cite{McInerney:2015xwa}, as many thermodynamic quantities can be written explicitly as relatively simple functions of $l$, $q$, $a$ and $X$. Additionally, $m$ with the new parameter $X$ facilitates analysis and also the area as a function of $X$ exhibits a symmetry that will be exploited in connecting KNdS thermodynamics with paramagnetic systems.
 Here we summarize the results that we use in later sections.\footnote{For more details, see Appendix \ref{appendix}.}
 
As is done in \cite{McInerney:2015xwa}, it is useful to define the parameter
 \be\label{xdef}
X=r_b r_c.
\ee
which is positive in the black hole parameter space. The sum of the areas of the black hole and de Sitter horizons,
\begin{align}
	A= A_b +A_c	,
\end{align}
takes a simple form in terms of the $X$ parameter: \footnote{ There is a typo the formula for $A$ in reference \cite{McInerney:2015xwa}, which is 
corrected here.}
\be\label{areax} 
A = 4\pi l^2 - {4\pi \over  \gamma} \left(X+\frac{\tilde{a}^2l^2}{X}\right),
\ee
where
\be\label{atdef}
 \at^ 2 = q^2 + a^2.
\ee
A key observation that we will investigate with the spin correspondence is that $A$ is symmetric under the transformation $X \rightarrow \tilde{a}^2l^2/X$:
\be 
A(X)=A(\tilde{a}^2l^2/X).
\label{eq:3}
\ee
\begin{figure}
\centering
\includegraphics[width=.45\linewidth]{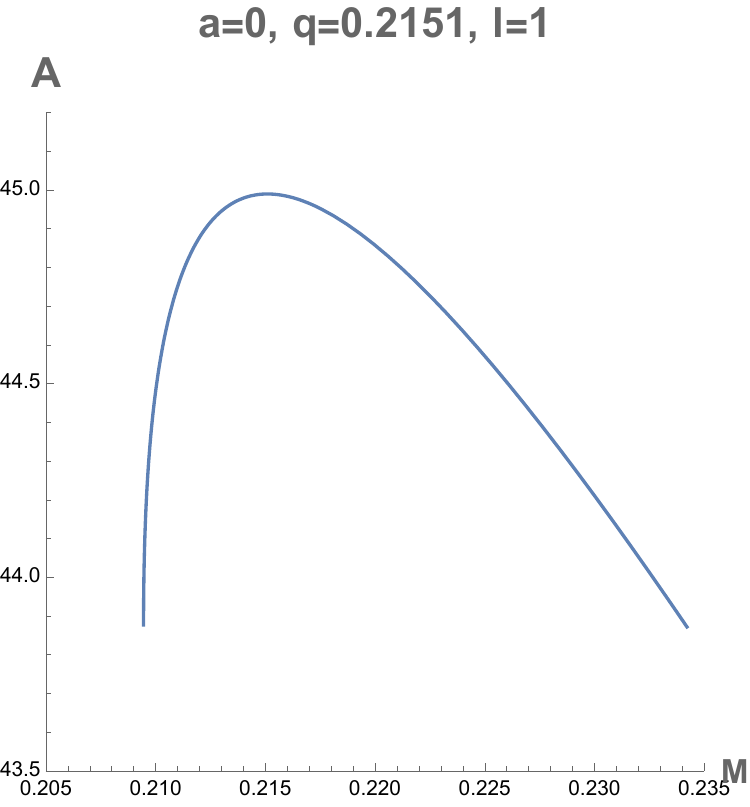}
\caption{Total area $A$ as a function of mass $M$. }
\label{f.AofM}
\end{figure}
This says that there are two distinct KNdS spacetimes (different values of $X$) with the same total area, with one proviso.
It is important to keep in mind that the transformation of $X$ can sometimes send $X$ out of the black hole parameter space \cite{McInerney:2015xwa}.
 When $X$ and $\tilde{a}^2l^2/X $ are both within the black hole parameter range, then there is necessarily an extremum of $A(X)$ 
between them and the symmetry (\ref{eq:3}) implies that this happens at the fixed point $X=\at l$. In fact, such an extremum is a maximum.

The metric parameter $m$ is given by
\be\label{massx} 
m={1\over 2l^2 } \left( l^2-a^2 + X-\tilde{a}^2l^2/X \right)^{1/2} \left(X+\frac{\tilde{a}^2l^2}{X}\right).
\ee
Formulae for the horizon radii and temperatures are given in the Appendix, see (\ref{qarad}), (\ref{khfactored}), and (\ref{khfactored2}).

\subsection{The $y$ parameter, and symmetry of the area and system surface gravity}

Motivated by the symmetry in equation (\ref{eq:3}), we introduce the variable $y$,
\be\label{ydef}
X=  \at l e^y, \qquad y= \ln{X\over \at l} .
\ee
Substituting into (\ref{areax}) gives
\be\label{areay}
A = 4\pi l^2 \left(1 -  {2 \at \over \gamma l} \cosh y \right )  ,
\ee
which is symmetric under $y\rightarrow -y$,
\be\label{areaysymm}
A(y) = A(-y).
\ee

This symmetry \eqref{areaysymm} displays the symmetry of the area in a form that is more analogous to the symmetry of the entropy
in the spin system studied in Section \ref{s.SpinSystems}. The $y$-parametrization also 
gives a simple way to calculate the ``system surface gravity" $\kappa_{\text{sys}}$, 
\be\label{ksdef}
\kappa_{\text{sys}}  =   8\pi {dM\over dA } 
\ee
This quantity will correspond to the temperature of the spins system. 
The metric parameter $m$ (\ref{massx}) as a function of $y$ is
\be\label{my}
m=  \at \cosh y \ h(y) \ , \quad h(y) = ( 1- {a^2 \over l^2} + 2 {\at \over l} \sinh y )^{1/2}. 
\ee
Taking the derivatives of the mass and of the area gives an effective surface gravity to the black hole system,
\be\label{ks}
\kappa_{\text{sys}}  =  - {1 \over l \gamma  } \left(  h(y) + {\at \over l} {\cosh^2 y \over h(y) \sinh y }\right),
\ee
where we have used that the ADM mass $M$ is related to the metric parameter $m$ by $M=m /\gamma^2$.
The quantity $\kappa_{\text{sys}}$ naturally arises in terms of $\kappa_b$ and $\kappa_c$ when studying the first laws for black holes in de Sitter, as we do below. It is often referred to as the effective temperature in the literature. The expression (\ref{ks}) is convenient because it gives  $\kappa_{\text{sys}}$
in terms of the four parameters $l$, $a$, $q$, and $y$. Since $h>0$,\footnote{  It follows from (\ref{alphapm} ), (\ref{xrange}), and (\ref{beta}) that $h(y)$ is always real and in the black hole parameter space. See  the
analysis in \cite{McInerney:2015xwa} for details.} it follows that $\ks$ can only vanish for $y<0$ which happens at the minimal mass limit.

Lastly, we note that using $y$ as a parameter facilitates writing $m$ as a function of $A$ using equations (\ref{areay}) and (\ref{my}). 
There are two branches depending on whether $y<0$ (smaller black holes) or $y>0$ (larger black holes),
\be\label{ma} 
m(A)=\frac{l}{2\gamma}\left( \gamma^2-\frac{A}{4\pi l^2} \right)\left[ 1-\frac{a^2}{l^2}\pm 2\left( \frac{1}{4\gamma^2}\left(\gamma^2-\frac{A}{4\pi l^2}\right)^2-\frac{\tilde{a}^2}{l^2} \right)^{1/2} \right]^{1/2}.
\ee
\begin{figure}
\centering
\includegraphics[width=.45\linewidth]{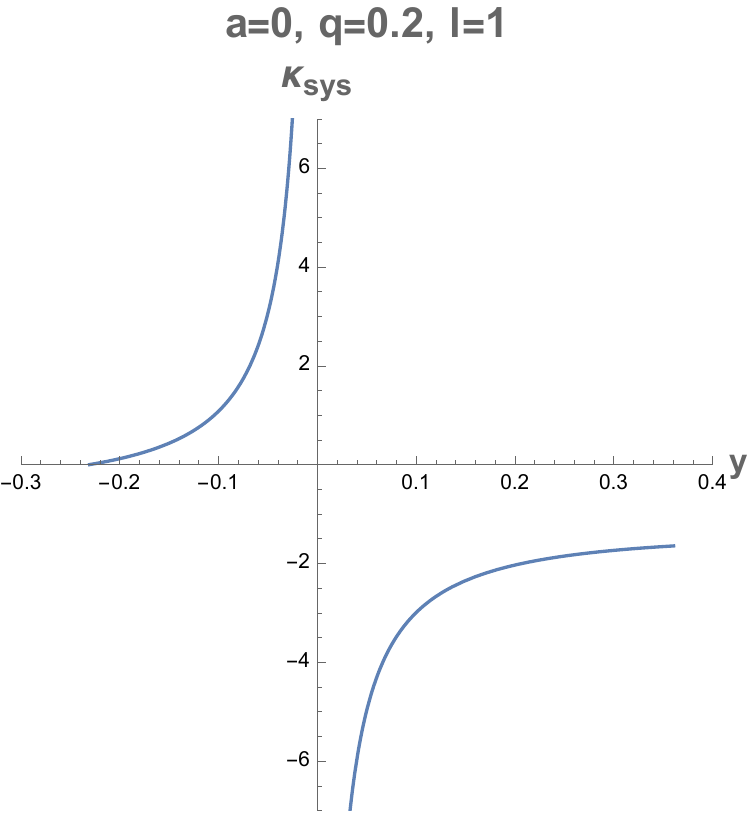}
\caption{Effective temperature $\kappa_\text{sys}$ as a function of $y$. }
\label{f.ksysofy}
\end{figure}

\subsection{Maximal entropy conditions and gravitational temperatures from the first laws}

In Schwarzschild de Sitter the maximal area spacetime with fixed $\Lambda = 3/l^2$ is pure de Sitter, corresponding to $M=0$ with area $A=4\pi l^2$. As the mass increases, the total area decreases
 monotonically until the black hole and cosmological horizons meet. When $r_b=r_c$ the mass has 
  its maximum value $M=l/(3\sqrt{3} )$, and the area reaches a minimum value
 $A= 8\pi l^2 /3$. 
 
 The total horizon area for 
rotating, charged black holes has a very different behavior from Schwarzschild de Sitter, for fixed $a$, $q$ and $\Lambda$. For small but non-zero values
 of $a$ and $q$, as the mass increases from $M_{\text{min}}$ to $ M_{\text{max}}$ then
 $A$ is maximized at some intermediate value of $M$.\footnote{ If the charge and angular momentum are allowed to go to zero, then
$A$ is maximized by pure de Sitter.}
As $a$ and $q$ increase, at some point $A(M) $ loses the interior maximum, and  increases monotonically from  $M_{\text{min}}$ to $ M_{\text{max}}$.
 These relations can be seen in Figure \ref{f.AofM}. The analytic conditions on $a/l$ and $q/l$  are worked out in \cite{McInerney:2015xwa}
 and summarized in the Appendix \ref{appendix}.

Using the first laws for KNdS we now derive
the conditions for maximal entropy and the thermodynamic interpretation of the conditions.
 The two first laws, one for the black hole horizon and one for the cosmological horizon, 
 are \cite{Dolan:2013ft}\footnote{  A relation of the same form holds for the inner horizon, but will not be used here.}

\begin{equation}\label{first}
\delta M = {\kappa_h \over 8\pi} \delta A_h +\Omega_h \delta J + \Phi_h \delta Q -V_h {\delta \Lambda \over 8\pi},
\qquad  h\in\{b,c\},
\end{equation}
where $\Phi_h = \phi_h - \phi_\infty$ is the difference between the value of the electric potential at the horizon $h$ and at infinity,
and $V_h$ is the thermodynamic volume. Alternatively, since $J=aM$, this can be written as
\begin{equation}\label{first2}
(1- a \Omega_h) \delta M = {\kappa_h \over 8\pi} \delta A_h +\Omega_h M\delta a+ \Phi_h \delta Q -V_h {\delta \Lambda \over 8\pi}
\qquad  h\in\{b,c\}.
\ee
The first form is useful when considering families of KNdS black holes at constant $J$, while the second form is especially
useful for families at constant $a$.

The first laws can be used to find  criteria for the maximal area spacetime for fixed $q$ and $\Lambda$, and
either fixed $J$ or fixed $a$, as follows. Setting $\delta J =0$, one can evaluate \eqref{first} at the two different horizons to obtain two equations, and then use these two equations to eliminate $M$ to obtain
\be\label{deltaA}
 \delta A=  \left ( 1 +{\kappa_b \over \kappa_c } \right) \delta A_b, \qquad \text{fixed $J$},
 \ee
 where $\delta  A= \delta A_b + \delta A_c $.
 As long as $\delta A_b$ is finite and non-zero, and $\kappa_c \neq 0$, then $\delta A =0$ when $ \kappa_b + \kappa_c =0$, \emph{i.e.},
\be\label{eqsg}
 \kappa_b = - \kappa_c  = | \kappa_c | \  .
 \ee
In fact, this extremum of $A(M)$ is a maximum, as can be seen in Figure \ref{f.AofM}.

Next, we find the maximal area condition for fixed $a$. Evaluating \eqref{first2} at the two horizons gives two formulas, and $\delta M$ can be eliminated to give
\be\label{deltaA2}
 \delta A= \left ( 1+ {\kappa_b \over \kappa_c} { 1- a\Omega_c \over 1- a\Omega_b } \right) \delta A_b. 
 \ee
 Hence, the condition for $\delta A =0$ is 
\be\label{eqsg2}
 \kappa_b   ( 1- a\Omega_c) = - \kappa_c  ( 1- a\Omega_b )  , \qquad \text{fixed }a,
 \ee
which differs from (\ref{eqsg}).

The two  conditions (\ref{eqsg}) and (\ref{eqsg2}) for a maximal area black hole make sense from the point of view of temperature in 
the microcanonical ensemble, in which the temperature of a subsystem $h$ is defined as
\be\label{tempdef}
{1\over T_{h, W } }=  \left({ \partial S_h \over \partial E_h} \right)_W,
\ee
where the derivative is taken at fixed work terms $W$. 
Applying this to the first laws (\ref{first}) gives the well known identification
\be\label{tempj}
T_{h,J}^{\text{grav}} = {\kappa_h \over 2\pi }  , \qquad \text{fixed }  J.
\ee
On the other hand, applying the definition to (\ref{first2}) gives
\be\label{tempa}
T_{h,a}^{\text{grav}}= {\kappa_h \over 2\pi ( 1- a\Omega_h )} , \qquad \text{fixed } a.
\ee
Hence, the different conditions for the maximal area black hole spacetime are both conditions that
 the magnitude of the temperatures are equal, but a different temperature is relevant in the two cases because
 different quantities are held fixed. 

For non-rotating uncharged black holes in de Sitter spacetimes, ``Schwarzschild de Sitter", $\kc$ is negative, and that continues to be true in KNdS (see equation (\ref{khfactored})).
 The fact that inner horizons and cosmological horizons have negative surface gravity has been an issue of some discussion in the literature.
 Reference \cite{Cvetic:2018dqf} shows that negative temperatures naturally arise in the Gibbsian formalism of thermodynamics.
 Reference \cite{Banihashemi:2022htw} resolves some questions raised by the negative value of $\kc$ by introducing a system boundary which allows for a better behaved thermodynamic ensemble. 
 In this paper, we will see that $\kc < 0$ is one of the needed ingredients 
 for KNdS thermodynamics to match that of the spin system.

For the metric (\ref{rotmetric}) equality of the magnitudes of the surface gravity requires  that
\begin{align}
	 - (\rbh ^2 + a^2 ) \Delta^\prime (\rds  ) = (\rds ^2 + a^2 ) \Delta^\prime (\rbh  ).
\end{align}
Solutions occur when either $\rds =\rbh$, which is when the two horizons coincide and the common temperature is zero, or at
$X^2 -2a^2 X -\at^2 l^2 =0$, with positive solution
\be\label{xeq}
X_{\text{eq}} = a^2 +\sqrt{ a^4 + \at^2 l^2 } , \qquad \text{fixed }J.
\ee

Turning to  $a=\text{constant}$, 
 the symmetry of the area (\ref{eq:3}) implies that $A$  is extremized at $X= \at l$. This value of $X$ is the solution to the equal temperature condition (\ref{eqsg2}). At this value,
\begin{align}\label{xeq2}
X_{\text{eq}} = \at l  ,\qquad  m_{\text{eq}} =\at \sqrt{1-a^2 /l^2 },\qquad
A =  4\pi l^2 \left( 1 - { 2 \at \over \gamma l}  \right).
\end{align}
When $a=0$ but $q$ is nonzero, the two equilibria agree, and this reduces to the
analysis of reference \cite{Romans:1991nq}.

It is interesting to speculate to what extent the terms in equations
  (\ref{tempj}) and (\ref{tempa}) can be viewed as temperatures. In these equations, the derivative is taken
with respect to the same energy $M$ for both $h=b$ and $h=c$, rather than with respect to an energy of the
subsystem $h$. The mass of the spacetime is not a local quantity and cannot be divided into a sum of two constituent masses. 
This issue does not arise if there is only one horizon.
We will return to this issue when considering subsystems of spin systems. 
 This criticism is also not relevant if 
 the variation of the total area $\delta A $ is related to $\delta M$. In this case, the coefficient of proportionality is
 often referred to as the effective temperature. A linear combination of the first laws (\ref{first}) gives
\be\label{dmtoda}
\delta M = {\kappa_b \kappa_c \over \kappa_b + \kappa_c } {\delta A \over 8\pi}, \qquad \text{fixed}\  J.
\ee
Whereas at fixed $a$, (\ref{first2}) gives
\be\label{dmtoda2}
\delta M = {\kappa_b \kappa_c \over (1-a\Omega_c ) \kappa_b + (1-a\Omega_b )  \kappa_c } {\delta A \over 8\pi}
, \qquad \text{fixed}\  a.
\ee
In keeping with using the term ``surface gravity" for gravitational quantities, and  ``temperature" for properties of 
spin systems, we will refer to the coefficient multiplying $\delta A/8\pi$ as the system surface gravity $\kappa_\text{sys}$.
For fixed $J$
\be\label{kappasys} 
\kappa_{\text{sys}} = {\kappa_b \kappa_c \over  \kappa_b +   \kappa_c }.
\ee
For fixed $a$
\be\label{kappasys2} 
\kappa_{\text{sys}} = {\kappa_b \kappa_c \over (1-a\Omega_c ) \kappa_b + (1-a\Omega_b )  \kappa_c }.
\ee

The system surface gravity can be positive or negative since $\kc$ is negative. 
For small black holes $\kappa_b \rightarrow 0$ and $\kappa_c$ is finite, so both of the expressions for $\ks$ go 
to zero through positive values in this limit. For large black holes both $\kb$ and $\kc$ both go to zero, so more information is needed
to take the limit. Equation (\ref{ks}) for $\ks$ shows that $\kappa_\text{sys}$
 goes to a negative constant in the small mass limit (see Figure \ref{f.ksysofy}) . We note that $\kappa_{\text{sys}}$ is a global quantity that depends on quantities intrinsic to both horizons. This is unlike the
 surface gravities $\kappa_h$ (and the angular velocities $\Omega_h$) which depend on only the geometry local to their horizons $h$.

 A key feature of both $\kappa_{\text{sys}}$ for constant $J$ or constant $a$ is that in each case $\kappa_\text{sys}$ diverges at the maximal area black hole,  
 which is seen by comparing the denominators of (\ref{kappasys}) and (\ref{kappasys2}) with  (\ref{eqsg}) and (\ref{eqsg2}). Further, at that point
 $\kappa_{\text{sys}}$ changes sign  from positive
to negative as the mass increases, due to the negative contribution of $\kc$. 
In the literature, $\ks / 2\pi$ is often referred to as the effective temperature, for example see the analysis in the review article
\cite{Kubiznak:2016qmn}.  This effective temperature switches from positive to negative, and diverges in between. 
Such a behavior  is characteristic of the temperature of 
 a spin system with a finite number of spins, which we turn to next.

 \section{Spin systems\label{s.SpinSystems}}
In this section we review the energy, entropy, and temperature of a non-interacting
system of spins and analyze the correspondence with KNdS black holes.

The behavior of the area of a rotating, charged black hole as a function of mass  in de Sitter spacetimes is
qualitatively different than that for a black hole in an asymptotically flat or AdS spacetime. This is due to the presence of the 
cosmological horizon which has its own surface gravity and
imposes a maximum area and a maximum mass of the black hole, whereas for $\Lambda \leq 0$ the black hole can be arbitrarily large.
In addition, $\kappa_b$ goes to zero when mass is minimized or maximized.  However, the total area and the system surface gravity are quite similar to the entropy and temperature of a two-state statistical  system with $N$ particles, such as a paramagnet, as we discuss next.  
   
Reference \cite{Dinsmore:2019elr} compared such spin systems to a Schwarzschild de Sitter spacetime, noting several features that 
are  qualitatively the same between the two systems, including the bounded mass/energy and area/entropy. 
The Schwarzschild black hole temperature $\kb $ goes to infinity as the mass goes to zero, but KNdS black holes are better behaved since $\kb$ 
equals zero when the mass is minimized. 

Some of the features for de Sitter black holes are also exhibited by paramagnets with finite degrees of freedom. In subsection \ref{uncs}, we consider a simple model involving a system of $N$ spins  interacting with a magnetic field $B$. This model reproduces some of the qualitative behavior for the KNdS black holes. In the subsection \ref{s.conssub}, we refine this model by considering two subsystems, which reproduces other properties of KNdS.

In this section, we begin by discussing an unconstrained spin system. This captures some of the qualitative features of KNdS black holes, but it misses the details of the individual horizons. To remedy this, we then discuss a spin system with two constrained subsystems that captures the qualitative features of the individual horizons.

\subsection{Unconstrained spin system}\label{uncs}

Consider a system of $N$ spins with magnetic moment $\sigma$ interacting with a magnetic field $B$. 
Let $N_\pm$ count the number of spins parallel and anti-parallel to the magnetic field, with $N_+ + N_- =N$.
The energy of a configuration of spins is 
\begin{equation}
 E= \sigma B (N_- -N_+)= \sigma B( 2 N_-  - N) 
\end{equation}
so $ -\sigma B N\leq E\leq \sigma B N$. 
In what proceeds, we shall set $N\gg 1$ and work in the microcanonical ensemble.  The number of microstates with fixed energy is  given by the binomial coefficient
\begin{equation}
\Omega(N_-) = {N\choose N_-} = {N!\over N_- !(N-N_-)!}.
\end{equation}
Define $\lambda$ as the ratio\footnote{The traditional statistical mechanics notation is $x=N_+ /N$, however we use $\lambda$ so as not to clash with the $X=r_br_c$ parameter in KNdS spacetimes.}   of the number of spins down (anti-aligned with the magnetic field) to the total number of spins,
\be\label{lambdadef}
\lambda =  {N_- \over N }.
\ee
In terms of this ratio, the energy of the system is
\be\label{energy}
E= \sigma BN (2\lambda -1 ) \ , \ \ 0\leq \lambda\leq 1.
\ee
From \eqref{energy}, $E$ increases with $\lambda$, just as $M$ increases with $X$ or $y$. 

For large $N$, the entropy is approximately
\begin{align}\label{entropy}
	S \simeq  N s( \lambda ) = &
N \left( -\lambda \log \lambda -(1-\lambda )\log (1-\lambda ) \right).
\end{align}
 The entropy is zero for $\lambda =0$ and $\lambda =1$, when all the spins are up or all down respectively,
  and has a maximum of $N\ln 2$ at $\lambda =1/2$ (see Figure \ref{f.spinfig.Slambda}). Since $\lambda$ and $E$ are related linearly, $y$ can be expressed as a linear function of $E$, and so $S$ as a function of $\lambda$ can be straightforwardly replaced with $S$ as a function of $E$. 
  
 From
  (\ref{entropy}), the entropy is symmetric under interchanging $\lambda$ and $1-\lambda $,
 \be\label{entropysym}
 s(\lambda )  = s( 1-\lambda ).
 \ee
 This corresponds to the fact that the counting  states does not distinguish between spins up or spins down.
 The energy of a spin, however, does depend on whether it is up or down. For this reason, as we will see, the entropy decreases in the high energy limit, and this system exhibits a Schottky anomaly in the entropy.
  
The temperature of the system is determined by the slope of the $S(E)$ curve,
\begin{equation}\label{spintempformula}
{1\over T}= {\partial S \over \partial E} = {1\over 2\sigma B}\log\left({1-\lambda \over \lambda }\right),
\end{equation}
and is plotted in Figure \ref{f.spinfig.Tlambda}.
 Because the entropy is maximized when the number of spins up equals the number of spins down, 
the slope, and hence the temperature, is positive for $0\leq \lambda < 1/2 $ and negative for $1/2 \leq \lambda \leq 1 $.
The temperature goes to infinity at the maximum of $S$.
There are two zero temperature limits. One is the lowest energy state where the temperature approaches zero through
positive values, and the other is the highest energy state,  which approaches zero temperature through negative values.

Negative temperatures occur when increasing the energy decreases the entropy. Here, this happens because increasing the energy increases the number of spins anti-aligned with the magnetic field, but when $\lambda>1/2$, the spins are mostly anti-aligned with the magnetic field, and this means that increasing the energy (making more spins anti-aligned) decreases the entropy. This negative temperature is consistent with the symmetry of $S$ as a function of $y$. The negative temperature states are unstable,
and under perturbations those states will tend towards increasing entropy while increasing their temperature to less negative
values.

\begin{figure}
\centering
\begin{subfigure}{.5\textwidth}
  \centering
  \includegraphics[width=.9\linewidth]{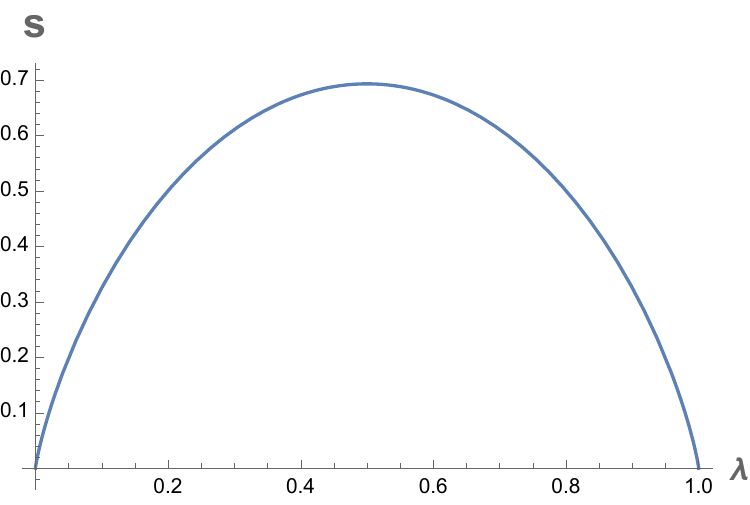}
  \caption{Entropy per spin, $s=S/N$, versus $\lambda$.}
  \label{f.spinfig.Slambda}
\end{subfigure}%
\begin{subfigure}{.5\textwidth}
  \centering
\includegraphics[width=.9\linewidth]{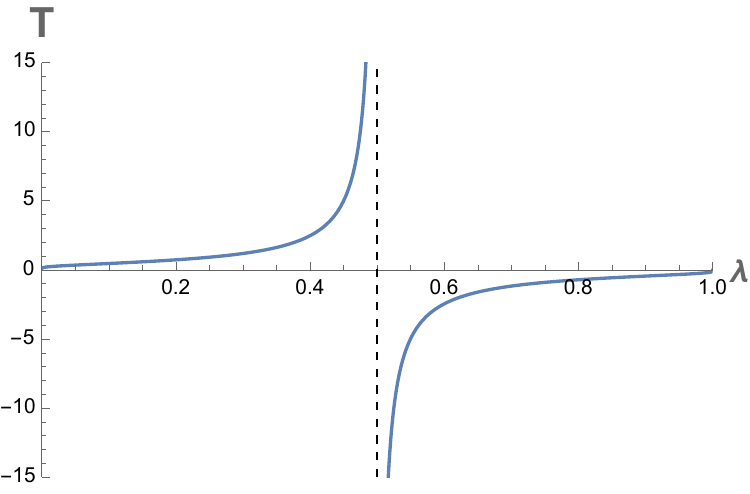}
  \caption{Temperature $T$ versus $\lambda$, with $2\sigma B=1$. }
  \label{f.spinfig.Tlambda}
\end{subfigure}
\caption{Entropy per spin and temperature plotted as functions of $\lambda=N_-/N$. }
\label{f.spinfig}
\end{figure}

Now compare the spin system to the gravitational system with the standard correspondence 
$E\sim M$, $S_h \sim A_h / 4$, and $2\pi T_h \sim \kappa_h  =  2\pi T_{h, J}^{\text{grav}}$ for fixed $J$, or  $2\pi T_h \sim (1- a\Omega_h )^{-1} \kappa_h  \equiv
T_{h, a}^{\text{grav}} $ for fixed $a$. 
The qualitative similarities between the behavior of  $S$ and $A$, and $T$ and $\kappa_{\text{sys}}$, are apparent when comparing Figure \ref{f.Aofy}
to Figure \ref{f.spinfig.Slambda}, and 
Figure \ref{f.ksysofy} to Figure \ref{f.spinfig.Tlambda}. The divergence in $\ks$ is seen to correspond to a divergence in $T$ in a well behaved
statistical system. Further, the negative values of $\ks$ are indicative of an underlying structure in which there are
a finite number of degrees of freedom, each with two states.

Particularly striking are the symmetries of $S(\lambda ) $ and $A( y) $,
 \be\label{symmcorresp}
\lambda \rightarrow 1-\lambda \  , \quad \text{and} \quad y\rightarrow -y .
\ee
(see Figures \ref{f.spinfig.Slambda} and \ref{f.Aofy} ).
The spin parameter $\lambda$ goes from zero to one, and $y$ can be linearly rescaled and shifted  to $y^\prime$
so that $0\leq y^\prime \leq 1$,
\begin{align}\label{yprime}
y^\prime & =  \log \left( {X\over X_{\text{min}} } \right) \left[ \log \left( {X_{\text{max}} \over X_{\text{min}} } \right)  \right ]^{-1}  = { \log( X/l^2 ) - \log (X_{\text{min}} /l^2 ) \over \log (X_{\text{max}} /l^2 )  - \log (X_{\text{min}} /l^2 ) } ,
\end{align}
where the factors of $l^2$ have been inserted to make the argument of the logarithm dimensionless.

We now compare $y^\prime$ to $\lambda = N_{-} / N $ expressed in terms of the energy,
\be\label{lambdae}
\lambda (E) = {E - E_{\text{min}} \over E_{\text{max}} - E_{\text{min}} }
\ee
where $E_{\text{max}/\text{min} } = \pm \sigma B N$.
The spin parameter $\lambda$ is the fraction of spins in their higher energy state, which increases as $E$ increases.
The corresponding quantity in (\ref{yprime}) is $\log X$, which is analogous to $E$
since $m$ varies monotonically with $X$. 
So the two relations (\ref{yprime}) and (\ref{lambdae}) 
suggest that $y^\prime$ is a measure of the fraction of gravitational constituents in their higher energy state, which
increases as  $\log X$ (or $m$) increases.

We conclude this section by relating $N$ to the KNdS parameters. 
The maximum of $A$ occurs at $ X_{eq} = \at l$ with value $\max (A) =  4\pi l^2 ( 1 -{ 2 \at / \gamma l })$.
Since the area
curve in the completely dual black hole case has the best qualitative match to the  entropy curve
of the unconstrained spin system (\ref{totdual}) we focus on that family. In that case  $ \sqrt{ {3\over 5}} \at ln \leq X  \leq  \sqrt{ {5\over 3}} \at l$, and 
$A$ has the same, minimal, value at each endpoint\cite{McInerney:2015xwa}. Hence
\be\label{mama}
{1\over 4 \hbar G }
\left( \max (A) - \min (A)\right) = {\at l \over 4 \hbar G \gamma }   \left( \sqrt{3/5 } + \sqrt{ 5/3 } - 2 \right) .
\ee
The numerical factor in parenthesis is approximately $0.07$. On this family of black holes, $q/ l$ and $ a/ l$ satisfy (\ref{totdual}) 
so  $q/ l , a/ l$  are less than one. In the spin system, 
\be\label{msms}
S_{\text{max}} - S_{\text{min}} = S_{\text{max}} = N \ln 2.
\ee
Equating (\ref{mama}) and  (\ref{msms}), substituting $l^2 =3/G\Lambda$, and $ \at =  \sqrt{ a^ 2 + q^2}$ gives a main resulting correspondence of this paper:
\be\label{e.Ncorresp}
N \sim  \  { \sqrt{ a^ 2 + q^2} \over \hbar G^{3/2} \Lambda^{1/2} }.
\ee
The mass does not appear in the expression for $N$. This is the result of comparing the maximum of $S$, which occurs at 
a particular value of $E$ to the maximum of $A$, which occurs at a particular value of $m$. 
Since $N$ is fixed its value does not change as $E$ changes.

\subsection{Refining with constrained subsystems}\label{s.conssub}
The above model reproduced some of the thermodynamic features KNdS black holes, in particular, the system temperatures and entropies, 
leading to a relation between $N$ and the black hole parameters given by \eqref{e.Ncorresp}.
However, this spin model is limited. It relates only totally self-dual black holes to spin systems. It says nothing about individual horizons in the black hole systems, and $T$ goes to zero as $\lambda \rightarrow 1$ whereas $\ks$ goes
to a negative constant.

The spin model can be refined by dividing the spin system into ``$b$"  and ``$c$" subsystems,  
each with their own entropy and temperature. 
Reference \cite{Dinsmore:2019elr} considered constrained states of the  paramagnet  model which was divided into $b$ and $c$ subsystems. Building on this, our next
goal is to find appropriate constraints on states so that the $b$ and $c$ subsystems qualitatively replicate the black hole
and horizon areas and surface gravities. 

In this refined model, suppose that the total number $N$ of spins is fixed, but now we subdivide the system into systems $b$ and $c$ with $N_b$ and $N_c$ spins, satisfying
\be\label{nbc}
N= N_b + N_c \ \ , \quad N_-  = N_{b-} + N_{c-},
\ee
where $N_{h-}$ is the number of aligned spins in the $h$ subsystem, and $N_-$ is the number of aligned spins in the entire system.
The states are labelled by $N_b$, $N_{b-}$, and $ N_{c-}$, $N_c $ is determined by $N_c =N - N_b $.
Define
\be\label{lambdah}
\lambda_h ={N_{h-} \over N_h } \ , \ \ \quad h\in\{b,c\},
\ee
so that states can alternatively be labelled by  $N_b, \lambda_b$ and $\lambda_c$. Then
the total entropy of the constrained system is
\begin{eqnarray}\label{sbc}
S_{\text{con}} &= & S_b + S_c  
= N_b s( \lambda _b )  + N_c s( \lambda _c ) .
\end{eqnarray} 
The maximum  possible value of $S_{\text{con}}$ is again $S_{\text{con}} = N\ln 2$
and occurs when $\lambda_b = \lambda_c =1/2$. However, when we impose additional constraints below to
match the physics of the spin systems to the KNdS black holes, the maximum value of $S_\text{con}$ can be lower than $N\ln 2$.

The total energy is 
\be\label{energybc}
E = E_b + E_c  
\ee
where $E_b$ and $E_c$ are the energies of the subsystems, given by
\be\label{eh}
E_h =\sigma B( N_{h-} - N_h ) = \sigma BN_h (2\lambda_h - 1 ).
\ee
The temperature of the $h$-th subsystem is
\begin{equation}\label{th}
{1\over T_h }= {\partial S_h \over \partial E_h } = {1\over 2 \sigma B}\log\left({1-\lambda_h \over \lambda_h }\right).
\end{equation}
 Since  $\kb>0$ and $\kc <0$, identifying $2\pi T_h$ with  $\kappa_h $ requires that $\lambda_h$ satisfies
\be
0\leq \lambda_b < {1\over 2} \  , \quad \text{and} \ \ {1\over 2}\leq  \lambda_c < 1.
\ee
Hence the naturally occurring negative temperature states of the spin system correspond to
the naturally occurring negative surface gravity of the cosmological horizon. 

An important difference between the black hole spacetime and the spin system is that there is no gravitational  analogue
of the division of energy between the two spin subsystems of \eqref{energybc}. This is because gravitational mass is not localized.
The same $\delta M$ appears in the first laws \eqref{first2} for the black hole and for the cosmological horizon because they are relations
between a boundary term at the black hole horizon or at the cosmological horizon respectively, and the \emph{same} boundary
at infinity. Therefore we need to select a constrained subset of all the spin states such that  $\delta E  = \delta E_b + \delta E_c =2 \delta E_b = 2\delta E_c $. Hence the correspondence requires that the subsystems satisfy the following first laws,
\begin{subequations}
\begin{align}\label{samede}
{1\over 2 }\delta E  &= \delta E_b =T_b \delta S_b  -\mu_b \delta N_b, \\
{1\over 2 }\delta E &= \delta E_c =T_c \delta S_c -\mu_c \delta N_c ,
\end{align}
\end{subequations}
with
\begin{align}
\delta N_b & + \delta N_c  =0,
\end{align}
where $\mu_h$ is the chemical potential for spin subsystem $h$. The last constraint is so that the total number of spins is fixed.

Dividing $\delta E_b$ and $\delta E_c$ in \eqref{samede} by $T_b$ and $T_c$, respectively, and then adding the results gives
\be\label{deds}
\delta E = T_{\text{sys}} \delta S_{\text{con}}  -\left( {\mu_b \over T_b } - {\mu_c \over T_c } \right) \delta N_b,
\ee
where
\be\label{tsys}
T_{\text{sys}} = \  { 2 T_b T_c \over T_b +T_c } .
\ee
The temperature of the constrained system $T_{\text{sys}}$ has the same form as $\ks$ (\ref{kappasys}) since the algebraic  forms of the 
first laws are the same. Both $T_{\text{sys}}$ and $\ks$ go to infinity when the temperatures or surface gravities have equal magnitudes. 
The difference in the factor of $2$ in the numerator of $T_{\text{sys}}$ arises from the definition in (\ref{energybc}).
Substituting (\ref{th}) gives
\be\label{tsys2}
T_{\text{sys}}= \  {4 \sigma B \over \log \left({1-\lambda_b \over \lambda_b }\right)
+ \log \left({1-\lambda_c \over \lambda_c }\right) }.
\ee

In what follows, for brevity of presentation, we drop the distinction between the gravitational temperatures (\ref{tempj}) and (\ref{tempa}), 
 which refer to fixed $J$ or fixed $a$ respectively, and simply adopt the correspondence $\kappa_h= 2\pi T_h$.\footnote{One can always 
 restore the factors of $(1-a \Omega_h )$, but it turns out that since $a/l$ is small, these are
 higher order corrections.}
 In principle, matching  a family of constrained spin states with a family of KNdS black holes could be done as follows.
Start by fixing $l$ and $B$.  A family of KNdS black holes, parameterized by $y$, is specified by picking values of $a/l$ and $q/l$ within the KNdS
parameter space. 
The curves $\kappa_h (y)$  are then specified by equation (\ref{khfactored}), and the curves $T_h (\lambda_h )$ are given in (\ref{th}). 
The relation between $\lambda_h$ and $y$ is determined by the equality
\be\label{lambdax}
T_h (\lambda_c  (y) ) =  \kappa_h  (y) 
\ee
and by starting the curves at the same value. To do this, note that
at $y=y_{\text{min}} $, $ \kappa_b= 0$ and $\kappa_c^{(0)} $ is a known non-zero value determined by 
the choice of $a$ and $q$. Setting $\lambda_b =0$ at the 
corresponding state makes $ T_b = 0$, and solving $2\pi T_c (\lcz ) =\kappa_c   (y_{\text{min}} ) $ for $\lcz (y_{\text{min}} ) $ gives the initial value
 $\lcz$.
 
The magnitude of the cosmological temperature decreases monotonically
to zero as $X$ goes from $y_\text{min}$ to $y_\text{max}$, so 
  $\lambda_c =1$ at $y=y_\text{max}$. On the other hand, the black hole temperature increases from zero at $y_\text{min}$ to a (finite) maximum  and then decreases to zero again at $y= y_\text{max}$.    Hence there are two branches to the solution 
giving $\lambda_b$ as a function of $y$, starting and ending at $\lambda_b =0$.  

Just as for the unconstrained spin system, a relation between $N$ 
and the KNdS parameters will be found by comparing the difference in the gravitational and spin entropies
at their maximal values minus their minimal values.

It remains unknown how $N_b$ might be, in principle, identified  with the KNdS parameters. To get
 $N_b ( y )$ one can write down a differential equation for $dN_b /dy$, since
  $\lambda_b (y)$ and $\lambda_c (y) $ are known implicitly, 
\be
 {1\over 4} {dA \over dy}  =  {dS_{\text{con
 }} \over d\lambda_b } {d\lambda_b \over dy} + {dS_{\text{con}} \over d\lambda_c } {d\lambda_c \over dy} 
 + {dS_{\text{con}} \over dN_b} {dN_b \over dy}.
 \ee
 Of course this set of calculations is algebraically challenging, in particular the first step of extracting  $\lambda_h (y)$ by setting the 
temperatures equal.
While solving this in general is difficult, in this paper we find the correspondence between the spin and gravitational systems
in the small and large mass limits, and at the maximal entropy points.

\subsection{Black hole and spin correspondence for constrained states in some limits}

We now compare the spin systems and the black holes in certain limits, and identify features of the correspondence between the KNdS and paramagnet parameters. As discussed above,
we identify the temperatures of the two systems, $ 2\pi T_h \sim  \kappa_h$ and $2\pi T_{\text{sys}}\sim \kappa_{\text{sys}}$.
For small mass black holes $\kb$ goes to zero but
 $\kc$ is not zero, so in the corresponding limit for the spins we require
  $\lambda_b \rightarrow 0$ and  $1/2<  \lambda^{(0)}_c < 1$.
In the maximal mass black hole limit,
$\kb\rightarrow 0^+ \ , \kc\rightarrow 0^-$, corresponding to 
 $\lambda_b \rightarrow 0$ and $\lambda_c \rightarrow 1$.

Equating the KNdS and spin temperatures leads to a mismatch in the values for the entropies, as $A_b$ and $A_c$ are nonzero 
 even when the surface gravities vanish, while  $S_h$  goes to zero as $T_h $ goes to zero. Thus, instead of setting
$A/4$ equal to $S$ we will set
  the differences between their maximum and minimum values equal to each other.

 \subsubsection{Small black holes}
 
 As $\lambda_b $ approaches zero, $S_b$ approaches zero. However $S_c$ is never zero,
 and so, when energy is minimized, the total entropy is
 \be\label{sconsmall}
 S_{\text{con}} \simeq S_c \simeq N^{(0)}_c s( \lcz ) , \qquad \text{$E$ minimized.}
 \ee
Small black holes correspond to a state with almost
 no $b$ spins down and a fraction $\lcz$ of $c$ spins down, since $\lambda_h = N_{h-} / N_h$. 
 The fraction $\lcz$ is determined by specifying $a, q$, finding $\kc$, and  then solving $T_c = T_{h,a}$
 given in equations (\ref{th}) and (\ref{tempa}) respectively. For the small black holes this is algebraically complicated  and will not be pursued here.

\subsubsection{ Large black holes}

The large black holes are characterized by the black hole and cosmological horizons approaching each other, so 
 $\kb\rightarrow 0^+ $ and  $\kc\rightarrow 0^-$ such that $\kappa_{\text{sys}}$ goes to a negative constant, see Figure \ref{f.ksysofy}.
In contrast,
$T$ for the unconstrained spins goes to zero in the high energy limit, see Figure \ref{f.spinfig.Tlambda}. 
The division into subsystems allows the specification of the rate at which each
 subsystem go to zero, so that $T_{\text{sys}}$ can approach a non-zero constant.
 To study this, let
 \be\label{largeflow}
 \lambda_b \simeq \mu  , \qquad \lambda_c \simeq 1-K \mu  ,\qquad \ K>0,
 \ee
with $ \mu \rightarrow 0$. Then $1/T_b \simeq -\log \mu / 2B$ and $1/T_c \simeq -\log (K \mu ) / 2B$, so
\be\label{largetemp}
T_{\text{sys}} \simeq {2B \over \log K }.
\ee
Having $T_\text{sys}$ be negative requires
\be\label{ksize}
K<1.
\ee
The particular value of $K$ is set by matching to the value of $\kappa_{\text{sys}} (X_{\text{max}} )$ for the choice of 
$a, q$. The entropies scale as
\be\label{largeentropies}
S_b \simeq -N_b \mu \log \mu \ \ , \quad S_c \simeq -N_c K \mu \log \mu .
\ee
In order that the entropies become equal, corresponding to $A_b \rightarrow A_c$, it follows that 
\be\label{kn}
N_b = K N_c \ <  N_c.
\ee
Thus, in this model the number of $b$ spins is less than 
the number of $c$ spins even when the areas are equal. Both $S_b$ and $S_c$ are going to zero, so
\be\label{sconbig}
  S_{\text{con}} \simeq 0  , \qquad \text{$E$ maximized.}
\ee
This limit corresponds to a state in which almost all  the $b$ spins are in their low energy state, and almost all the
 $c$ spins are up in their high energy state,
with both subsystems approaching zero temperature. 

Hence the curve $S_{\text{con}}(E) $ is no longer symmetrical.  
 On the lower energy side of the curve
all states have a dual state with the same
entropy, while higher energy states with entropies lower than $N^{(0)}_c s( \lcz )$ do not
have a dual state. The corresponding
black holes were called  ``small dual"  in the classification of \cite{McInerney:2015xwa} because only the smaller
mass black holes have duals. These small dual black holes occur in the portion of
the $a/l , q/l$ parameter space bounded 
by the ``totally dual" family (\ref{totdual}) and extends to arbitrarily small
but non-zero $ a/l $ and $ q/l$ so that there is still an inner horizon where $\kappa_b$ goes to zero.

\begin{figure}
\centering
\begin{subfigure}{.5\textwidth}
\includegraphics[width=.9\linewidth]{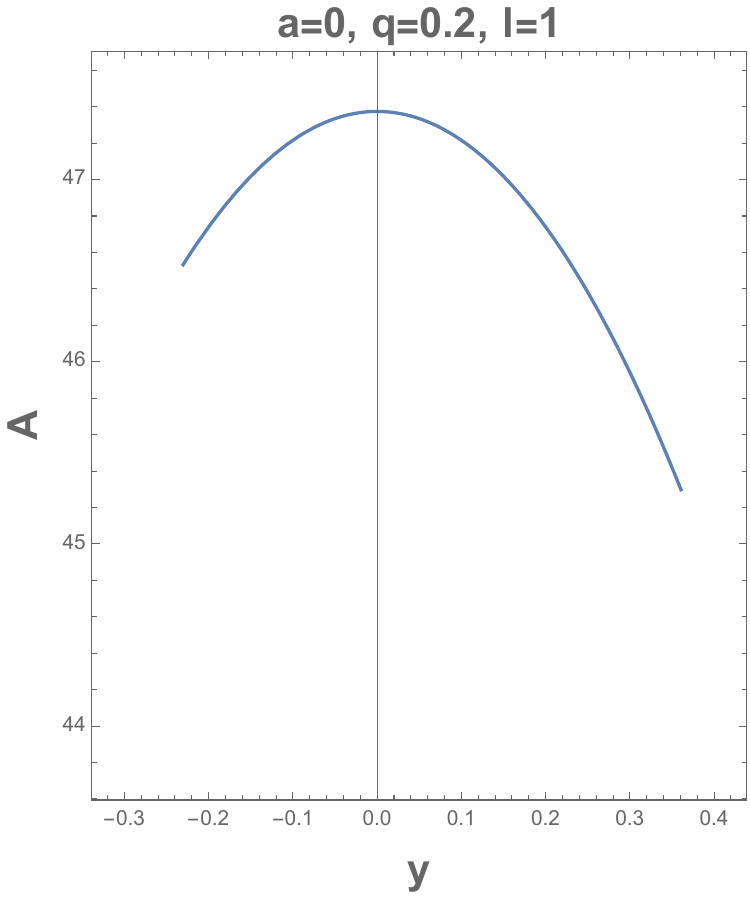}
 \caption{Small dual case}
  \label{f.yASD}
\end{subfigure}%
\begin{subfigure}{.5\textwidth}\includegraphics[width=.9\linewidth]{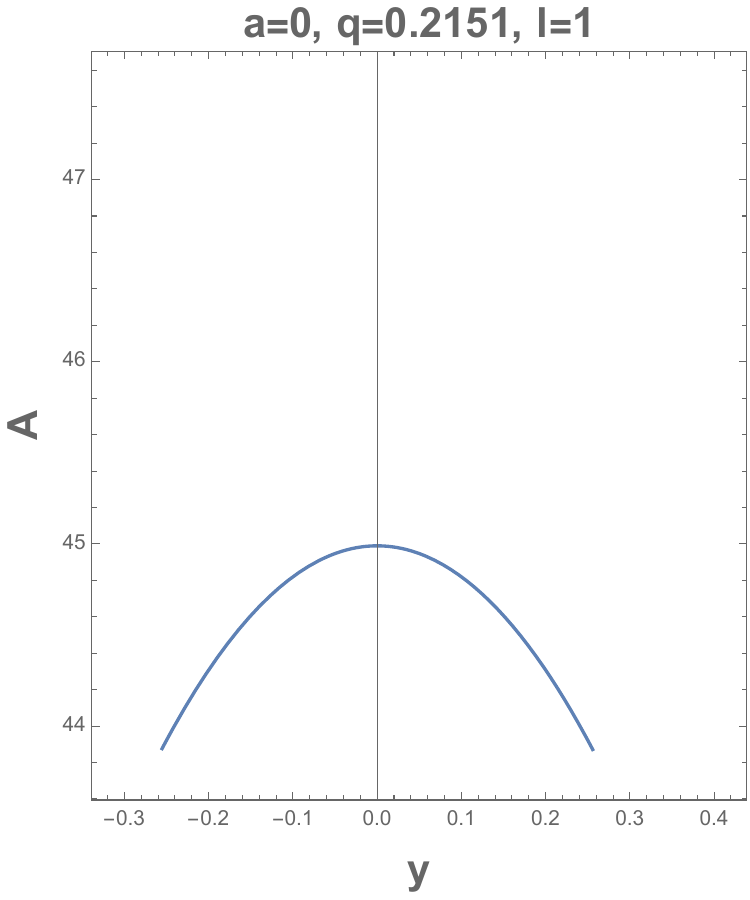}
  \caption{Totally dual case}
  \label{f.yATD}
\end{subfigure}
\caption{Total area $A$ as a function of $y$}
\label{f.Aofy}
\end{figure}

\begin{figure}
\centering
\begin{subfigure}{.5\textwidth}
  \centering
  \includegraphics[width=.9\linewidth]{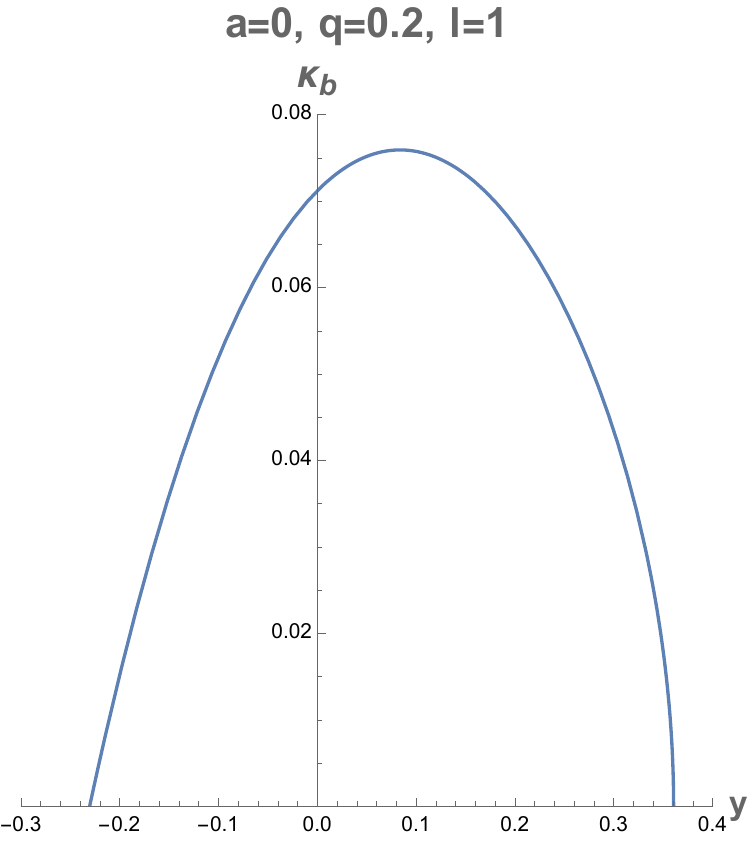}
  \caption{$\kappa_b$ as a function of $y$ for small-dual black holes.}
  \label{f.ykappabSD}
\end{subfigure}%
\begin{subfigure}{.5\textwidth}
  \centering
\includegraphics[width=.9\linewidth]{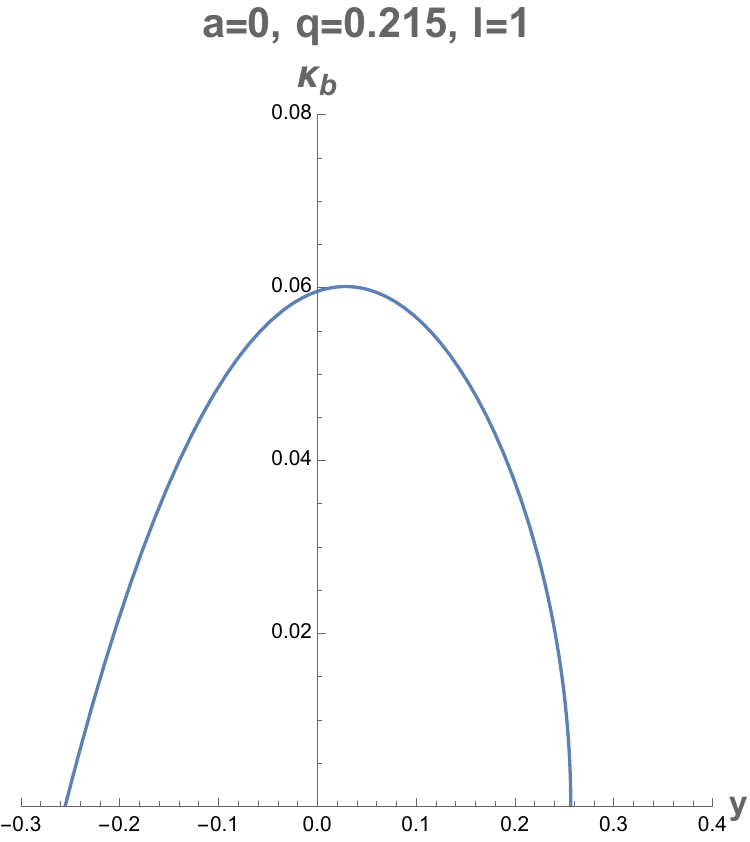}
  \caption{$\kappa_b$ as a function of $y$ for totally-dual black holes.}
  \label{f.ykappabTD}
\end{subfigure}
\begin{subfigure}{.5\textwidth}
  \centering
  \includegraphics[width=.9\linewidth]{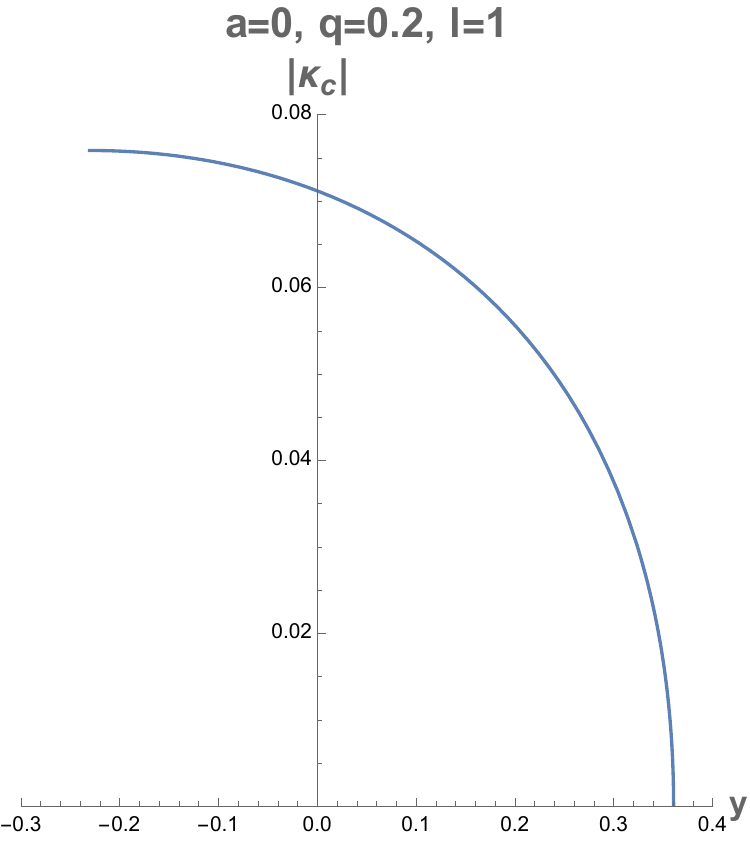}
  \caption{$|\kappa_c|$ as a function of $y$ for small-dual black holes.}
  \label{f.ykappacSD}
\end{subfigure}%
\begin{subfigure}{.5\textwidth}
  \centering
\includegraphics[width=.9\linewidth]{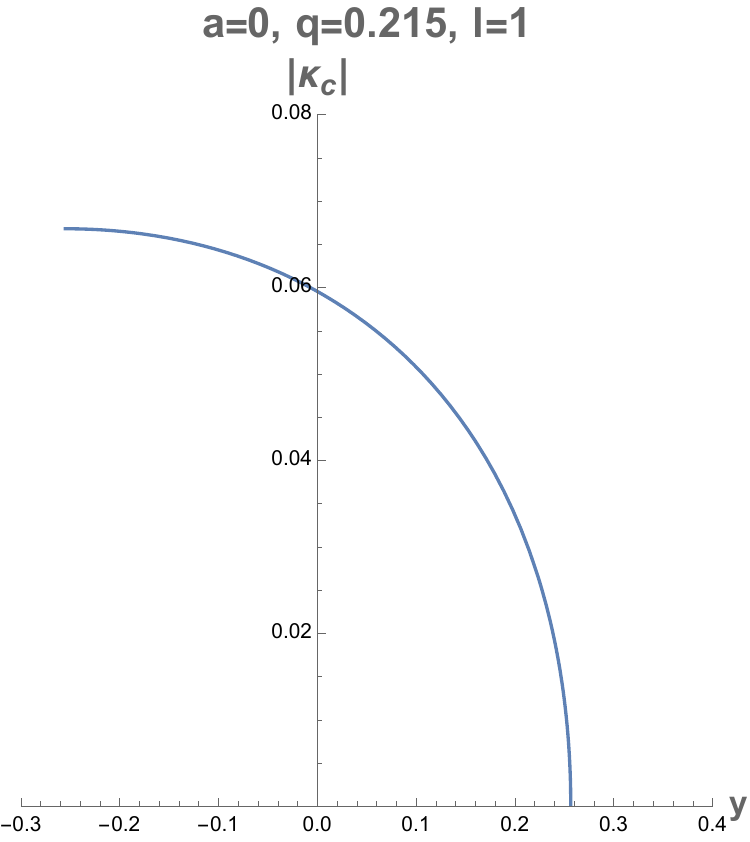}
  \caption{$|\kappa_c|$ as a function of $y$ for totally-dual black holes.}
  \label{f.ykappacTD}
\end{subfigure}
\caption{$\kappa_b$ and $|\kappa_c|$ as functions of $y$ for small-dual and totally-dual black holes.}
\label{f.ykappa}
\end{figure}

A natural question is whether the spin model contains a family of substates corresponding to the ``large dual" black holes of
 \cite{McInerney:2015xwa}, and thus covering the full KNdS parameter space. A family of large dual 
 black holes is one in which all the large black holes have a smaller mass dual, but some small black holes
 do not.  The answer appears to be no if the boundary
 conditions on $T_b$ and $T_c$ are kept, since this enforces $S_{con} \simeq 0$ at the large black hole end.

\subsubsection{Maximal entropy: equal magnitude temperatures}

Comparing the maximal entropy of the spin system with the black holes area gives the relation (\ref{e.Ncorresp}) between
$N$ and $\at l$ for the totally dual black holes. These black holes have $a/l, q/l$ on the curve (\ref{totdual}) in the KNdS parameter space.

We proceed analogously with the constrained system. Because of the behavior of the entropy (\ref{sconbig}) the constrained
subsystems allow us to find $N$ over the small dual portion of the KNdS parameter space.
The main result is given in (\ref{e.Ncorresp3}). Some of the algebra has been moved to the Appendix (\ref{appsg}).

In both the black hole and the spin systems, the maximal entropy state 
is where the $b$ and $c$ temperatures have equal magnitude, and so
 $\kappa_{\text{sys}} $ and $T_{\text{sys}}$ go to infinity. For the black holes, this happens at $X=\at l$.
For the spins, this happens when $\lambda_b = \lambda_b^{\text{eq}} = 1-\lambda_c^{\text{eq}} $. To find $\lambda_b^\text{eq}$, we need to find
$\kappa_b^{\text{eq}} $ at $X=\at l$, and then solve for $\lambda_b^{\text{eq}}$ by equating the gravitational and spin temperatures.
 The expressions for the surface gravities
(\ref{khfactored})  are algebraically complicated to work with due to the $X$ dependence of the $r_h$. 
However the product
of the two gravitational temperatures depends on $X$ in a simpler way, and when they have equal magnitudes the product is
(minus) the square of the temperature at equilibrium. Since we are considering KNdS families at constant $a, q$, the spin 
temperature is to be equated with the gravitational temperature (\ref{tempa}), rather than $\kappa_h / 2\pi$.
The strategy then is to compute both sides of
\be\label{eqsgeqt}
  {\kappa_b \kappa_c \over 2\pi ( 1- a\Omega_b ) ( 1- a\Omega_c )} = 4\pi ^2 T_b (  \lambda_b^{\text{eq}} )T_c (1- \lambda_b^{\text{eq}} ),
 \ee
which gives $\lambda_b^{\text{eq}}$ as a function of $a$ and $q$. Then, $S_{\text{con}} (\lambda_b^{\text{eq}} )$ can be compared to
$A( \at l )$ to get a relation between $N$ and $a/l , q/ l$ as was done in  (\ref{e.Ncorresp}).

Some details of the calculation of the product of the surface gravities are given in the appendix. Since $a/l $ and $q/l$
are constrained to be small (see (\ref{paramsp})), we give only the leading order result here,
\be\label{sgprod}
 - {\kappa_b^{\text{eq}} \kappa_c^{\text{eq}} \over 2\pi ( 1- a\Omega_b ) ( 1- a\Omega_c )} \simeq {3\over 4} {1\over \at l}.
 \ee
 It turns out that to this order in $\at l$, $1- a\Omega_b \approx 1$,
 so the distinction between the two gravitational temperatures for constant $a$ or constant $J$ is irrelevant.
 
 The  product of the spin temperatures at equal  magnitude temperatures  is 
 \be\label{spintprod}
 -T_b (\lambda_b^{\text{eq}} ) T_c ( 1-\lambda_b^{\text{eq}} )  = ( 2 \sigma B)^2 \left( \log {1- \lambda_b^{\text{eq}} \over \lambda_b^{\text{eq}} } \right)^{-2}.
 \ee
Substituting (\ref{sgprod}) and (\ref{spintprod})  into (\ref{eqsgeqt}) then gives
\be\label{lbeq}
\lambda_b^{\text{eq}} = {1\over   e^\rho +1  }\ \  , \quad \quad \rho ={2\sigma B l \over\sqrt{3} \pi} \left( { \at  \over l}\right)^{1/2}.
\ee
Hence $\lambda_b^{\text{eq}} \rightarrow 1/2$  when $\rho$ goes to zero and $ \lambda_b^{\text{eq}} \rightarrow 0$ when $\rho$ goes to infinity,
so $0< \lambda_b^{\text{eq}} < 1/2$ (as it should be for the $b$ spins).

The maximum entropy is thus
\begin{align}\label{maxscon}
\begin{aligned}
	S_{\text{con}}^{\text{eq}}   & = N_b s (\lambda_b^{\text{eq}} ) + N_c s (1- \lambda_b^{\text{eq}} ) \\
& = (N_b + N_c ) s(\lambda_b^{\text{eq}} ) \\
 &= N\left( {\ln (1+ e^\rho ) \over 1+ e^\rho } + {\ln (1+ e^{-\rho}  ) \over 1+ e^{-\rho }  } \right) .
\end{aligned}
\end{align}
where the relation \eqref{lbeq} between $\lambda_b$ and $\lambda_c$ at equilibrium, and the symmetry \eqref{entropysym} of $s(\lambda )$, have been used. 
From \eqref{maxscon}, $S_{\text{con}}/N$ is maximized at  $S_\text{con}/N=\ln 2$ when $\rho =0$. 
The value of $\lambda_b^{\text{eq}}$ is determined by $\sigma Bl$ as well as $\at /l$. 
This is the first expression that prompts us to compare the values of $\sigma B$ and $l=\sqrt{3/G\Lambda} $. Both
$B$ and $\Lambda$ function as background fields in the analyses, $\sigma$ determines the strength of the coupling of the 
spins to $B$, and $G$ determines the strength of the coupling of the metric to $\Lambda$. 
 We proceed with the simple assumption that
\be\label{bl}	
\sigma B  \sim \sqrt{G \Lambda }\  ,
\ee
which implies that $\rho\lesssim 1$. If $\at / l \ll1$, then $\rho \ll 1$ and
  $\lambda_b^{\text{eq}} \simeq {1\over 2} (1- {\rho \over 2})$, which gives for the maximum entropy
  \be\label{maxscon3}
   S_{\text{con}}^{\text{eq}}    \simeq N \ln 2,
  \ee
just as in the unconstrained case.

 As $\at /l$, and so $\rho$, increase from close to zero to
 $\rho \sim \mathcal{O} (1)$, the constrained spin system corresponds to the small dual black holes, which
 can be seen as follows. The boundary of the small dual portion of the parameter space is given by (\ref{totdual}), so $\at /l$ is always small. Following the nomenclature of \cite{Susskind:2021dfc}, when 
  $A_b \ll A_c $ we call the black holes parametrically small, and when $A_b$ is of order $A_c$, 
  parametrically large.
  Small dual KNdS black holes are characterized by $A(y_\text{min}) \geq A(y_\text{max} )$ with equality at the 
 totally dual curve (\ref{totdual}). 
 The qualitative features of $A$ and $\kappa_c$ in the small dual portion of the KNdS phase space are shown in Figures \ref{f.Aofy} and \ref{f.ykappa}. 
 
For fixed $\at / l$, $|\kappa_c |$  monotonically decreases from $y_\text{min}$ to zero at $y_\text{max}$. 
 As $\at / l$ increases, the magnitude $|\kappa_c^{(0)} |$ at $y_\text{min} $ decreases.  
  Turning to the spin system, the prescription to match $\kappa_c$ with $T_c$ then
 implies that for fixed $\rho$, we have ${1\over 2} <\lcz < \lambda_c^{\text{eq}} < 1 $, and that 
$ \lcz $ increases as $\rho  $ increases. This causes an overall
decrease in the function $s(\lambda_c )$  and the entropy at the low energy end $s(\lcz )$ decreases most rapidly.
The entropy at the high energy end is fixed at zero. For the black holes, 
  as $\at/l $ increases, $A (y)$ 
 decreases pointwise such that the value of  $A$ at $y_\text{min} $ decreases faster than
  the value at  $y_\text{max}$, so that $A(y_\text{min})$ approaches $ A(y_\text{max})$.
   
 So the behavior of $S_{\text{con}}$ qualitatively matches the behavior
of the plots of $A$ as $\at /l$ increases. One difference is that since $\kc (y_\text{min} ) $ is non-zero, then  $\lcz < 1$, and
$S_c (\lcz ) $ is non-zero, so the curves for $S$ get close to, but  do not actually reach, the totally dual black hole curve for $A$.

We conclude that the small dual portion of the KNdS parameter space can be included 
in comparing $A$ to $S_{\text{con}}$. Let us see how that impacts the estimate
for $N$ previously derived for the unconstrained system  (\ref{e.Ncorresp}). 
At $\rho =1$, the value of the entropy is $S_{\text{con}} = N F(1) \simeq 0.63 N $ which is quite close to $\ln 2 N \simeq 0.69 N$.
So in comparing $S_{\text{con}}$ to the gravitational entropy we simply use $S_{\text{con}}\sim \ln 2 N$.
Hence the maximal value of the entropy for the constrained spin system is approximately the same as
for the unconstrained system.

The specifics of the small dual parameter space
have been worked out in \cite{McInerney:2015xwa}, finding that the minimum and maximum values of $X$
are in the ranges  $ \sqrt{ 3/5} \   \at l < X_{\text{min}} <  \at l $, and
$ \sqrt{ 5/3} \ \at l < X_{\text{max}} < \beta $, where $\beta$ is given in (\ref{beta}). So
the value of $X_{\text{max}}$ 
varies from $\sqrt{ 5/3 }\ \at l $, which is the totally dual black hole boundary,
 to approximately $l^2 / 3$ when $\at /l \ll 1 $, see
(\ref{betasmalla}) and (\ref{betabiga}).
 The minimum of $A$  occurs for the largest black hole at $X= X_{\text{max}}$.  In all cases the maximum of $A$ is at $X_{\text{eq}} =\at l$.
Subtracting,
\be\label{mama2}
{1\over 4 \hbar G }
\left( \max (A) - \min (A)\right) =    {\pi l^2 \over 3 \hbar  G  \gamma} \left( 1- 2{ \at^2 \over l^2} \right) , \qquad {\at \over l} \ll 1
\ee
Equating (\ref{mama2}) to the  maximum entropy of the spin system  gives for the parametrically small black holes
\be\label{e.Ncorresp2}
N\simeq {\pi l^2 \over 3 \hbar G  } \left( 1-2 {\at \over l} \right)  , \qquad {\at \over l} \ll 1,
\ee
where factors of order one have been dropped. 

The estimate for $N$ for the parametrically large black holes 
is given by (\ref{e.Ncorresp}). 
Assembling the results we find that over the range of $a/l , q/ l$ extending from  almost de Sitter to totally dual black holes,
 $N$ corresponds to
\be\label{e.Ncorresp3}
N:\    {1 \over \hbar G^2  \Lambda } \ \ \rightarrow \ \  {   \sqrt{ a^ 2 + q^2} \over \hbar G^{3/2} \Lambda^{1/2}  }.
\ee
When the black hole entropy is negligible, $N$ is approximately $A_c$ in Planck units. In de Sitter, this is the only
dimensionless combination. 
 However, with a black hole there are additional dimensionful parameters.
Equation (\ref{e.Ncorresp3}) gives the interpolation between de Sitter and  a black hole with $A_b$ of order $A_c$, 
 and is one of the main results of this paper.

\section{ Conclusions}

In this paper, we have shown that simple paramagnetic systems exhibit many of the same thermodynamical features of KNdS black holes. The total horizon area as a function of mass $A(M) $ of KNdS
 spacetimes
is qualitatively similar to the entropy as a function of energy $S(E)$ for a non-interacting system of $N$ spins in a magnetic field.
It follows that the system surface gravity $\kappa_\text{sys}$ and $T$ are also qualitatively similar. Both $M$ and $E$ are bounded.
$A$ and $S$ each have a maximum at a value of mass/energy intermediate between the smallest and largest values, so
 $\kappa_\text{sys}$ and $T$ are positive on the smaller mass/energy side, and negative on the larger side.
These features of $A(M)$ and $S(E)$ are imposed by a symmetry in each system. For the spins the symmetry is that the number of states 
with $n$ spins up is the same as the number of states with $n$ spins down, but that the energy of the two configurations differs.
 For the black holes, $A$ has a symmetry implying that there are two KNdS solutions with the same $A$, but having different mass. We have seen that when the algebraic form of the two symmetries is made to match, their physical interpretations also qualitatively match, see
equations (\ref{yprime}) and (\ref{lambdae}). 

If a small amount of energy $\delta E > 0$ is added to these systems, the black hole and spin entropies change  by
\be\label{achange}
\delta A = {\delta E \over 2\pi \ks } , \qquad \quad \delta S = {\delta E \over T }.
\ee
The change is positive or negative depending on whether $\ks$ or $ T$ is positive or negative.
 The best match to the unconstrained spin system are 
the totally dual KNdS black holes, since the energy/mass as a function of entropy is double-valued.
 The fact that the entropy can sometimes increase and sometimes decrease is a feature of temperature sometimes being positive and sometimes negative, which can happen in systems with bounded degrees of freedom, such as the spin system we have described in this paper.

The KNdS spacetime has distinct black hole and cosmological horizons, with $\kb$ positive and $\kc$ negative. This motivates dividing the
spin system into $b$ and $c$ subsystems, where  $T_b$ is constrained to be positive and $T_c$ negative.   
The area/entropy of each subsystem changes in response to an addition of energy $\delta E$ according to 
first laws analogous to (\ref{achange}).

In this paper, we have found constraints
on the divided spin system so that it better qualitatively reproduces the behavior of KNdS black hole thermodynamics. We are partially successful in this. In particular, we are able to match the constrained spin system
with a larger part of the KNdS parameter space. By comparing the gravitational and the spin entropies, we derive an expression
for the number of spins $N$ in terms of the KNdS parameters $a$, $q$, $l$ in (\ref{e.Ncorresp3}).
However, we have not explicitly solved all of the equations for the constrained spin system, and doing so could lead to 
better insights.

There are several other directions for future research.
KNdS spacetimes display a rich set of phenomena that occur in different parts of their parameter space. For instance, there are Schottky anomalies that appear or disappear at various loci in the parameter space. A complete charting of this space and these Schottky anomalies is the subject of future research. 
Including the area of the inner horizon in an analysis would require  
a more complicated spin system involving three subsystems with additional constraints. It would be interesting to understand
how the analysis changes if two-state spins were replaced with spins that have a larger number of states, and these last two
points may be connected. Lastly,
adding interactions between the spins will be crucial for reproducing the non-linear effects of the gravitational system, and for studying 
the instability implicit in negative temperatures.

\section*{Acknowledgements} We are greteful to David Kastor for many useful discussions. ME received support from NSF grants PHY-1914934 and PHY-2112800. YQ 
was supported by the National Key R\&D Program of China (2021YFC2203100), NSFC under Grant No. 12275146, and the Dushi Program and Shuimu Fellowship of Tsinghua University.

\appendix
\section{Thermodynamic review of KNdS\label{appendix}}
\subsection{More thermodynamic quantities for KNdS} \label{s.KNdSthermo.PS}
This appendix contains formulas for KNdS thermodynamic quantities.
The formulas in this paper before (\ref{totdual}) appear in \cite{McInerney:2015xwa}. 
The horizon radii $r_h$ occur when $\Delta (r_h ) =0$, with $\Delta$ given by (\ref{delta}). Solving for $m$ gives
\begin{equation}\label{zeros}
2m = - {r_h ^3 \over l^2 }  + r_h \left(1 - {a^2 \over l^2 }\right ) +{\qt^2 \over r_h } , \qquad h\in\{b,c\}.
\end{equation}
Computing the surface gravities $2 \kappa_h = (r_h^2 +a ^2) ^{-1} \Delta^\prime (r_h )$ and using (\ref{zeros}) to eliminate $m$ gives
\be\label{kndstemps}
2(r_{h}^2+a^2) \kappa_{h} =
  \left( \left((1- {a^2 \over l^2} \right)r_{h}  -3{ r_{h}^3 \over l^2 }  -{\at^2 \over r_{h} } \right). 
\ee 
Using the definition 
$ X=r_b r_c$ and $L= r_c -r_b$, $L$ can be expressed in terms of $X$,
\begin{equation}\label{givesl}
L^2 = l^2 -a^2 -3X - {\qt^2 l^2 \over X},
\end{equation}
and the horizon radii are given in terms of $ X$  by 
\begin{equation}\label{qarad}
r_{b/c}   =\mp {L\over 2} + {1\over 2} \sqrt{ L^2 + 4X } ,
\end{equation}
where the minus sign is for $r_b$ and the plus sign for $r_c$. 

The expressions for $A(X)$ and $m(X)$  are given above in (\ref{areax}) and (\ref{massx}).
The surface gravities are given in terms of $X$ using (\ref{kndstemps}) and (\ref{qarad}). Another useful form is
\begin{equation}\label{khfactored}
 \kappa_h =
 - {3\over 2 l^2  } { (r_{h}^2- \alpha_- )(r_{h}^2-\alpha_+ ) \over r_h  (r_{h}^2+a^2) },
\end{equation} 
where $\alpha_\pm$ are the zeros of $L^2 (X)$,
\begin{equation}\label{alphapm} 
 \alpha_\pm = {l^2\over 6} \left(  \left(1-{a^2 \over l^2 } \right) \pm
 \sqrt{ \left( 1- {a^2 \over l^2 } \right)^2 - 12  {\at^2 \over l^2  }  } \right).
\end{equation} 
For processes at constant $a$ rather than $J$, as discussed above, the gravitational temperatures are $2\pi \kappa_h /(1- a\Omega_h )$,
given by
\begin{equation}\label{khfactored2}
 \kappa_h =
 - {3\over 2 l^2  } { (r_{h}^2- \alpha_- )(r_{h}^2-\alpha_+ ) \over r_h  (r_{h}^2 - a^4 /l^2) }.
\end{equation} 

\subsection{Black hole parameter space}

For the metric (\ref{rotmetric}) to describe a black hole in a spacetime with a cosmological horizon, the metric function $\Delta (r) $
must vanish at $r_b$ with slope greater than or equal to zero ($\kappa_b \geq 0$) and vanish
 at $r_c$ with slope less than or equal to zero ($\kappa_c \leq 0$). Hence
the physical parameter space is bounded by two geometries where $\kappa_b =0$, which becomes
\be\label{xrange}
\beta \leq  X \leq \alpha_+
\ee
where
\be\label{beta}
\beta = -\alpha_- + \sqrt{\alpha_- ^2 + \at^2l^2}.
\ee
Reality of $\alpha$ and $\beta$ then gives the following condition on the $a/l, q/l$ parameter space:
\be\label{paramsp}
 0 \leq \left( 1- {a^2 \over l^2} \right)^2 -12 {\at^2 \over l^2 } = 1 -12{q^2 \over l^2} - 14{a^2 \over l^2 } +{a^4 \over l^4}.
\ee
$a/l$ and $ q/l$ can be arbitrarily small, but one of them has to be non-zero to produce an inner horizon and enforce the
condition $\kb \rightarrow 0$ at the smallest, but non-zero, area. 
The magnitudes of $a/l$ and $q/l$ are relatively small; for $a= 0$, $q/l < 1/ \sqrt{12}$, and for 
 $q= 0$, $a/l + a^2 / (\sqrt{12} l^2 )  < 1/ \sqrt{12}$.
\be\label{betasmalla}
\beta \simeq {l^2 \over 3}\left( 1-4{a^2 \over l^2 } -3{q^2 \over l^2 } \right)  \ , \quad {\at^2 \over l^2 } \ll 1
\ee
\be\label{betabiga}
\beta \simeq {l^2 \over 6}\left( 1-{a^2 \over l^2 } -3{q^2 \over l^2 }\right) , \qquad  \text{largest $ a$, $q$.}
\ee

The totally dual family in which every spacetime has a dual with the same entropy is defined by $\alpha_+ = \at l/ \beta$,
which gives 
\be\label{totdual}
{108\over 5} {q^2\over l^2 } +{118\over 5} {a^2\over l^2 } - {a^4\over  l^4 }   =1.
\ee
Black holes with values of $a/l, q/l$ enclosed in this ellipse are called ``small dual" 
black holes, because all the smaller mass ones are dual to a spacetime with the same total area, but the largest mass
black holes are not. Note that  $a$ and $q$ can be arbitrarily small, but at least one must be nonzero to have an
inner horizon.

\subsection{Product of surface gravities}\label{appsg}

The expressions for $\kappa_h$ are cumbersome because the $r_h$ are functions of $X$ (\ref{qarad}).
However the product of the surface gravities can be written as a ratio of polynomials in $X$,
\be\label{sgprod2}
 -{ \kappa_b \kappa_c \over (1- a\Omega_b ) (1- a\Omega_c ) }  = {9\over 4 l^2 } {  {\cal N } \over {\cal D}} 
 \ee
 where
 \begin{subequations}
\begin{align}\label{sgprod3}
& {\cal N} = X^4 + X^2 ( \alpha_+^2 + \alpha_-^2  )  + \alpha_+^2 \alpha_-^2 + 
\alpha_+ \alpha_- ( r_b ^2 + r_c^2 )^2  - (  \alpha_+^2 \alpha_- +  \alpha_+ \alpha_-^2  )( r_b ^2 + r_c^2 )  \\
&   r_b ^2 + r_c^2 = l^2 - a^2 -X -\at^2 l^2 /X \\
& {\cal D} =  X \left[ X^2 -{a^4 \over l^2 } \left((l^2 -a^2 -X - { \at^2 l^2 \over X }\right)  + {a^8 \over l^4 } \right].
\end{align}
 \end{subequations}
The needed combinations of $\alpha_\pm$ are
\be\label{alphas}
\alpha_+ + \alpha_-  ={1 \over 3}(l^2- a^2 ) \   , \quad\quad 
\alpha_+  \alpha_ = {1\over 3} \at^2 l^2 
\ee
\be
\alpha_+^2 + \alpha_-^2 = 
{1\over 9} \left[ \left(1- {a^2 \over l^2} \right)^2 - 6\at^2 l^2 \right] .
\ee

\bibliographystyle{JHEP}
\bibliography{KNdSrefs}

\end{document}